\documentclass[preprint,preprintnumbers,nofootinbib]{revtex4-1}
\usepackage{graphicx}% Include figure files
\usepackage{dcolumn}% Align table columns on decimal point
\usepackage{bm}% bold math
\usepackage{amsmath}
\usepackage{amsfonts} 
\usepackage{latexsym}
\usepackage{bbm}
\usepackage{color}
\usepackage{amssymb}
\usepackage{amsthm}
\usepackage{epsfig}
\usepackage{physics}
\usepackage{hyperref}
\hypersetup{%
setpagesize=false,
 bookmarksnumbered=true,%
 bookmarksopen=true,%
 colorlinks=true,%
 linkcolor=blue,
 citecolor=red,
}

\begin{document}

\preprint{\textbf{OCHA-PP-371}}

\title{CP-violating effects on gravitational waves in a complex singlet extension of the Standard Model with degenerate scalars
}

\author{Gi-Chol Cho$^1$}
\email{cho.gichol@ocha.ac.jp}
\author{Chikako Idegawa$^2$}
\email{c.idegawa@hep.phys.ocha.ac.jp}
\author{Eibun Senaha$^{3,4}$}
\email{eibunsenaha@vlu.edu.vn}
\affiliation{$^1$Department of Physics, Ochanomizu University, Tokyo 112-8610, Japan}
\affiliation{$^2$Graduate school of Humanities and Sciences, Ochanomizu University, Tokyo 112-8610, Japan}
\affiliation{$^3$Subatomic Physics Research Group, Science and Technology Advanced Institute, Van Lang University, Ho Chi Minh City, Vietnam}
\affiliation{$^4$Faculty of Applied Technology, School of Engineering and Technology, Van Lang University, Ho Chi Minh City, Vietnam}
\bigskip

\date{\today}

\begin{abstract}

We examine CP-violating effects on electroweak phase transition (EWPT) in the standard model with a complex singlet scalar focusing particularly on a scenario where additional scalars have masses close to 125 GeV.   
Such a high mass degeneracy makes collider signatures in the scenario standard model like, and current experimental data cannot distinguish them from the standard model predictions.
We utilize a simplified scalar potential to understand impacts of CP violation on EWPT qualitatively. Then, one-loop effective potential with a thermal resummation is employed for full numerical evaluations. 
As a phenomenological consequence, gravitational waves from the first-order EWPT are also evaluated. 
We find that the strength of the first-order EWPT would get weaker as the CP-violating effect becomes larger. 
As a result, gravitational wave amplitudes are diminished by the size of the CP violation. 
Future gravitational wave experiments may shed light on CP violation in the singlet scalar sector as well as the experimental blind spot due to the high mass degeneracy.
\end{abstract}

\maketitle

%%%%%%%%%%%%%%%%%%%%%%%%%%%%%%%%%%%%%%%%%%%%%%%%%%%%%
% Introduction
%%%%%%%%%%%%%%%%%%%%%%%%%%%%%%%%%%%%%%%%%%%%%%%%%%%%%
\section{Introduction}\label{sec:intro}
The existence of physics beyond the standard model (SM) is strongly suggested by cosmological observations such as dark matter (DM)
and baryon asymmetry of the Universe (BAU). Despite this tangible experimental evidence, their relevant energy scales are still unknown, and diverse scenarios have been proposed. Extended models with new particle masses around $\mathcal{O}(100-1000)$ GeV are of particular interest from the viewpoint of high-energy physics experiments. 
Dedicated new physics searches by Large Hadron Collider(LHC)~\cite{ATLAS:2019nkf,CMS:2018uag} and DM direct detection experiments~\cite{Aprile:2020vtw} have made room for new physics at this scale narrowed to a large extent. It might indicate that some suppression mechanism or alignment should be at work if new particles are still there. 
The so-called degenerate scalar scenario can account for such an experimental blind spot. A representative example is an extension of the SM with a complex singlet scalar field $S$ (so-called CxSM~\cite{Barger:2008jx}, see also Refs.~\cite{Barger:2010yn,Gonderinger:2012rd,Coimbra:2013qq,Costa:2014qga,Costa:2015llh,Jiang:2015cwa,Chiang:2017nmu,Muhlleitner:2017dkd,Cheng:2018ajh,Grzadkowski:2018nbc,Chen:2019ebq,Basler:2020nrq,Abe:2021nih,Cho:2021itv,Egle:2022wmq} ) in which a mass of the additional scalar is close to 125 GeV~\cite{Abe:2021nih}.
The highly degenerated mass spectrum can yield a suppressed spin-independent cross section of DM ($\text{Im}S$) with nucleons,
moreover, the mass mixing angle between the degenerate scalars can be maximal without conflicting with LHC data~\cite{Khachatryan:2014ira}.

In Ref.~\cite{Cho:2021itv}, the current authors studied DM and first-order electroweak phase transition (EWPT) required by electroweak baryogenesis (EWBG)~\cite{Kuzmin:1985mm} (for reviews, see, e.g., Refs.~\cite{Quiros:1994dr,*Rubakov:1996vz,*Funakubo:1996dw,*Riotto:1998bt,*Trodden:1998ym,*Bernreuther:2002uj,*Cline:2006ts,*Morrissey:2012db,*Konstandin:2013caa,*Senaha:2020mop}) in the CxSM taking the degenerate scalar scenario. 
In contrast to the previous study~\cite{Abe:2021nih}, the suppression mechanism for the DM cross-section off nucleons is not compatible with the requirement of the strong first-order EWPT.  
Nevertheless, we can still find viable parameter space, but the DM mass is limited to around 62.5 GeV and 2 TeV. The 2 TeV DM is just on the border that is allowed by the DM relic density and direct detection data. Away from this point is excluded by either of them except for the 62.5 GeV case, where the DM annihilation cross section is enhanced resonantly, contributing to only tiny portion of the observed DM relic density and evading the DM direct detection constraints.

In the CxSM, the stability of DM is guaranteed by CP conservation of the Higgs sector, i.e., $\text{Re}S$ and $\text{Im}S$ do not mix. Given the fact that EWBG calls for CP violation beyond the SM and 62.5 GeV DM is far from being main component of the observed DM relic density, it is our next step to consider CP-violating (CPV) Higgs sector in the CxSM~\cite{Coimbra:2013qq,Costa:2014qga,Costa:2015llh,Muhlleitner:2017dkd} by demoting $\text{Im}S$ to an ordinary decaying particle. 

It is known that the CPV phase in this model does not generate a pseudoscalar coupling between the $\text{SU(2)}_L$ doublet Higgs and SM fermions, nor does CPV current. Therefore, additional new particles, especially fermions that couple to $S$, would be required to transmit the CPV effect to the SM matter sector and drive BAU. 
Nonetheless, as far as the dynamics of EWPT is concerned, scalars play a leading role, and the analysis within the CxSM is expected to give useful guidance towards more complete analysis. 

In this paper, we quantify the impacts of CPV on first-order EWPT, focusing particularly on the degenerate scalar case, which remains attractive as the experimental blind spot regardless of DM physics. To clarify differences from the previous CP-conserving (CPC) case~\cite{Cho:2021itv} qualitatively, we analyze EWPT using tree-level potential with thermal masses (referred to as high-temperature (HT) potential in this work). On the other hand, a finite temperature one-loop effective potential with Parwani resummation method~\cite{Parwani:1991gq} is used for a full numerical study. 
As a phenomenological consequence of the first-order EWPT, we also evaluate gravitational waves (GWs)~\cite{Witten:1984rs,Hogan:1986qda}~induced by the dynamics of bubbles and thermal plasma, which may provide a complementary probe of the experimental blind spot.

The paper is organized as follows. In Sec.~\ref{sec:model}, we introduce the CxSM and set our notation and input parameters. In Sec.~\ref{sec:ewpt}, we present how to calculate EWPT and bubble nucleation. In addition, the qualitative study of EWPT is also demonstrated using the HT potential. All the formulas necessary for the GW calculation are given in Sec.~\ref{sec:gw}. 
Sec.~\ref{sec:sum} is devoted to conclusion and discussions.

%%%%%%%%%%%%%%%%%%%%%%%%%%%%%%%%%%%%%%%%%%%%%%%%%%%%%
%Model
%%%%%%%%%%%%%%%%%%%%%%%%%%%%%%%%%%%%%%%%%%%%%%%%%%%%%

\section{Model}\label{sec:model}
The CxSM is the extension of the SM by adding the complex $\text{SU(2)}_L$ gauge singlet scalar field $S$.
The model was proposed in Ref.~\cite{Barger:2008jx} and shows that DM can exist if the scalar potential is invariant under CP transformation $S\to S^*$, and a vacuum expectation value (VEV) of $S$ is real. Phenomenological studies of DM can be conducted in Refs.~\cite{Barger:2010yn,Gonderinger:2012rd}.
Furthermore, one could obtain the strong first-order EWPT in this model ~\cite{Basler:2020nrq,Jiang:2015cwa,Chiang:2017nmu,Cheng:2018ajh,Grzadkowski:2018nbc,Chen:2019ebq,Cho:2021itv}.
 
The most general CPC scalar potential contains 11 new parameters~\cite{Barger:2008jx}. Since not all of them are relevant to DM and EWPT physics, we work on a minimal scalar potential defined as
\begin{align}
V_{0}(H, S)=\frac{m^{2}}{2} H^{\dagger} H+\frac{\lambda}{4}\left(H^{\dagger} H\right)^{2}+\frac{\delta_{2}}{2} H^{\dagger} H|S|^{2}+\frac{b_{2}}{2}|S|^{2}+\frac{d_{2}}{4}|S|^{4}+\left(a_{1} S+\frac{b_{1}}{4} S^{2}+\text {H.c.}\right).
\label{V0}
\end{align}\\
where both $a_1$ and $b_1$ break a global U(1) symmetry, avoiding an unwanted massless particle, and $a_1$ is introduced to avoid a domain wall problem which can arise when $Z_2$ symmetry $V_0(H, S)\to V_0(H,-S)$ is spontaneously broken.\footnote{We could consider other $Z_2$ breaking terms such as $S^3$ or $H^\dagger HS$ to avoid the domain wall problem. However, we exclusively focus on the potential Eq.~(\ref{V0}) by reason of the minimality and to make a comparison with our previous work~\cite{Cho:2021itv}, where Eq.~(\ref{V0}) is considered in the CP-conserving limit. } If the Higgs sector is CP conserving, real and imaginary parts of $S$ do not mix with each other and the latter can be the DM candidate. As mentioned in Sec.~\ref{sec:intro}, however, EWBG needs new CPV, and moreover, a condition of the strong first-order EWPT in this minimal CxSM renders the DM relic density much smaller than the observed value in the lower viable DM mass window $\sim 62.5$ GeV~\cite{Cho:2021itv}, requiring additional stable particle as a main component of the DM abundance. It is therefore more reasonable to consider CxSM including CPV and leave the DM issue to a more complete theory.

We parametrize the scalar fields as
\begin{align}
H(x) &=\left(\begin{array}{c}
G^{+}(x) \\
\frac{1}{\sqrt{2}}\left(v+h(x)+i G^{0}(x)\right)
\end{array}\right), \\
S(x) &=\frac{1}{\sqrt{2}}\left(v_{S}^{r}+i v_{S}^{i}+s(x)+i \chi(x)\right) \nonumber\\
& = \frac{1}{\sqrt{2}}\left(|v_S|e^{i\theta_S}+s(x)+i \chi(x)\right)
\end{align}
where $h$ is the SM-like Higgs field and $v(\simeq 246.22~\text{GeV})$ is its VEV, while $G^0$ and $G^+$ are Nambu-Goldsone (NG) fields. $v_S^r$ and $v_S^i$ are the VEVs of $s$ and $\chi$, respectively. In the CPC limit, $\chi$ becomes DM.
 For later use, we define $a_1=a_1^r+i a_1^i$ and $b_1=b_1^r+i b_1^i$.

At the tree level, tadpole conditions with respect to $h$, $s$ and $\chi$ are, respectively, given by
\begin{align}
&\left\langle\frac{\partial V_{0}}{\partial h}\right\rangle=v\left[\frac{m^{2}}{2}+\frac{\lambda}{4} v^{2}+\frac{\delta_{2}}{4}\left|v_{S}\right|^{2}\right]=0,\label{dvdh} \\
&\left\langle\frac{\partial V_{0}}{\partial s}\right\rangle=v_{S}^{r}\left[\frac{b_{2}}{2}+\frac{\delta_{2}}{4} v^{2}+\frac{d_{2}}{4}\left|v_{S}\right|^{2}+\frac{b_{1}^{r}}{2}\right]+\sqrt{2} a_{1}^{r}-\frac{1}{2}b_1^iv_S^i=0, \label{dvds} \\
&\left\langle\frac{\partial V_{0}}{\partial \chi}\right\rangle=v_{S}^{i}\left[\frac{b_{2}}{2}+\frac{\delta_{2}}{4} v^{2}+\frac{d_{2}}{4}\left|v_{S}\right|^{2}-\frac{b_1^r}{2}\right]-\sqrt{2} a_{1}^{i}-\frac{1}{2}b_1^iv_S^r=0,\label{dvdchi}
\end{align}
where $\langle\cdots \rangle$ denotes that all fluctuation fields are taken zero after the derivative. 

Note that $v_S^r$ is nonzero if $a_1^r\neq0$, which is important when discussing phase transitions in this model.
We also note that even though $a_1\neq0$, one could encounter another domain wall in the case of spontaneous CPV (called CP domain wall)~\cite{Chen:2020wvu} with $a_1^i = b_1^i = 0$. In this case, $V_0$ is invariant under $Z_2$ transformation $\chi \to -\chi$. Once the $Z_2$ symmetry is broken spontaneously, the CP domain wall would appear. After straightforward calculation, one finds
\begin{align}
v_S^i = \pm \sqrt{-v_S^{r2}+\frac{2\lambda}{\delta_2^2-\lambda d_2}
\left(-\frac{\delta_2m^2}{\lambda}+b_2+\frac{\sqrt{2}a_1^r}{v_S^r}\right)}.
\end{align}
However, this vacuum degeneracy is resolved when the explicit CPV is present, making the CP domain wall unstable.
We, therefore, assume explicit CPV throughout this study.\footnote{
According to Ref.~\cite{Chen:2020wvu}, explicit CPV with $\mathcal{O}(10^{-24}-10^{-23})$ in magnitude would be enough to avoid the big-bang nucleosynthesis constraints. 
GW signatures of the domain wall collapses are studied in detail there.}

The tree-level masses of the scalars are obtained by
\begin{align}
-\mathcal{L}_\text{mass}
& =
 \frac{1}{2}\left(\begin{array}{lll}
h & s & \chi
\end{array}\right) \mathcal{M}_{S}^{2}\left(\begin{array}{c}
h \\
s \\
\chi
\end{array}\right)
=
\frac{1}{2}\left(\begin{array}{lll}
h_1 & h_2 & h_3
\end{array}\right) 
O^{T} \mathcal{M}_{S}^{2} O
\left(\begin{array}{c}
h_1 \\
h_2 \\
h_3
\end{array}\right)
=\frac{1}{2}\sum_{i=1}^3m_{h_i}^2h_i^2,
\label{MM}
\end{align}
with
\begin{align}
 \mathcal{M}_{S}^{2} = 
\begin{pmatrix}
\frac{\lambda}{2}v^2 & \frac{\delta_2}{2}vv_S^r & \frac{\delta_2}{2}vv_S^i \\
\frac{\delta_2}{2}vv_S^r  & \frac{d_2}{2}v_S^{r2}-\frac{\sqrt{2}a_1^r}{v_S^r}+\frac{b_1^i}{2}\frac{v_S^i}{v_S^r} & -\frac{b_1^i}{2}+\frac{d_2}{2}v_S^rv_S^i \\
 \frac{\delta_2}{2}vv_S^i & -\frac{b_1^i}{2}+\frac{d_2}{2}v_S^rv_S^i & \frac{d_2}{2}v_S^{i^2}+\frac{\sqrt{2}a_1^i}{v_S^i}+\frac{b_1^i}{2}\frac{v_S^r}{v_S^i}.
\end{pmatrix},
\end{align}
where $m^2$, $b_2$, and $b_1^r$ are eliminated using the tadpole conditions 
\begin{align}
m^{2} &=-\frac{\lambda}{2} v^{2}-\frac{\delta_{2}}{2}\left|v_{S}\right|^{2}, \\
b_{2} &=-\frac{\delta_{2}}{2} v^{2}-\frac{d_{2}}{2}\left|v_{S}\right|^{2}-\sqrt{2}\left(\frac{a_{1}^{r}}{v_{S}^{r}}-\frac{a_{1}^{i}}{v_{S}^{i}}\right), \\
b_{1}^{r} &=-\sqrt{2}\left(\frac{a_{1}^{r}}{v_{S}^{r}}+\frac{a_{1}^{i}}{v_{S}^{i}}\right).\label{b1r}
\end{align}
The mixing matrix $O$ is parametrized as
\begin{align}
O(\alpha_i)
&=\left(\begin{array}{ccc}
1 & 0 & 0 \\
0 & c_{3} & -s_{3} \\
0 & s_{3} & c_{3}
\end{array}\right)\left(\begin{array}{ccc}
c_{2} & 0 & -s_{2} \\
0 & 1 & 0 \\
s_{2} & 0 & c_{2}
\end{array}\right)\left(\begin{array}{ccc}
c_{1} & -s_{1} & 0 \\
s_{1} & c_{1} & 0 \\
0 & 0 & 1
\end{array}\right),
%&=\left(\begin{array}{ccc}
%c_{1} c_{2} & -s_{1} c_{2} & -s_{2} \\
%s_{1} c_{3}-c_{1} s_{2} s_{3} & c_{1} c_{3}+s_{1} s_{2} s_{3} & -c_{2} s_{3} \\
%s_{1} s_{3}+c_{1} s_{2} c_{3} & c_{1} s_{3}-s_{1} s_{2} c_{3} & c_{2} c_{3}
%\end{array}\right),
\end{align}
where $s_i=\sin\alpha_i$ and $c_i=\cos\alpha_i~(i=1,2,3)$. \par
Let us summarize our input parameters. 
There are 9 degrees of freedom in the scalar potential, $\{m^2$, $\lambda$, $\delta_2$, $b_2$, $d_2$, $a_1^r$, $a_1^i$, $b_1^r$, $b_1^i\}$, where $\{m^2$, $b_2$, $b_1^r\}$ are traded with 3 scalar VEVs. Without loss of generality, we take $b_1^i = 0$. The remaining 5 degrees of freedom are fixed by $\{m_{h_1}$, $m_{h_2}$, $m_{h_3}$, $\alpha_1$, $\alpha_2\}$. In this work, we set $m_{h_1}=125$ GeV.

Here we list relationships between the input parameters and original Lagrangian parameters. 
From Eq.~(\ref{MM}), it follows that 
\begin{align}
\left(\mathcal{M}_{S}^{2}\right)_{i j}=\sum_{k} O_{i k} O_{j k} m_{h_{k}}^{2},
\label{Mij}
\end{align}
from which one finds
\begin{align}
\lambda &=\frac{2}{v^{2}} \sum_{i} O_{1 i}^{2} m_{h_{i}}^{2}, \\
\delta_{2} &=\frac{2}{v v_{S}^{r}} \sum_{i} O_{1 i} O_{2 i} m_{h_{i}}^{2}=\frac{2}{v v_{S}^{i}} \sum_{i} O_{1 i} O_{3 i} m_{h_{i}}^{2}, \label{del2}\\
d_{2} &=\frac{2}{v_{S}^{r 2}}\left[\frac{\sqrt{2} a_{1}^{r}}{v_{S}^{r}}+\sum_{i} O_{2 i}^{2} m_{h_{i}}^{2}\right]=\frac{2}{v_{S}^{i 2}}\left[-\frac{\sqrt{2} a_{1}^{i}}{v_{S}^{i}}+\sum_{i} O_{3 i}^{2} m_{h_{i}}^{2}\right]=\frac{2}{v_{S}^{r} v_{S}^{i}}\left[\sum_{i} O_{2 i} O_{3 i} m_{h_{i}}^{2}\right].
\label{d2fromMM}
\end{align}
Furthermore, the expressions of $d_2$ in Eq.~(\ref{d2fromMM}) yield
\begin{align}
a_{1}^{r}&=-\frac{v_{S}^{r}}{\sqrt{2}}\left[\sum_{i} O_{2 i}\left(O_{2 i}-O_{3 i} \frac{v_{S}^{r}}{v_{S}^{i}}\right) m_{h_{i}}^{2}\right],\\
a_{1}^{i}&=\frac{v_{S}^{i}}{\sqrt{2}}\left[\sum_{i} O_{3 i}\left(O_{3 i}-O_{2 i} \frac{v_{S}^{i}}{v_{S}^{r}}\right) m_{h_{i}}^{2}\right].
\end{align}
Note that $\alpha_3$ is not an independent parameter and determined by the second equality in Eq.~(\ref{del2})
\begin{align}
\sum_{i} O_{1 i}\left[\frac{O_{2 i}}{v_{S}^{r}}-\frac{O_{3 i}}{v_{S}^{i}}\right] m_{h_{i}}^{2}=\frac{(\mathcal{M}_S^2)_{12}}{v_S^r}-\frac{(\mathcal{M}_S^2)_{13}}{v_S^i}=0.\label{alp3}
\end{align}

In this model, Higgs couplings to fermions ($f$) and gauge bosons ($V=W^\pm,Z$) are, respectively, given by
\begin{align}
\mathcal{L}_{h_{i} \bar{f} f} &=-\frac{m_{f}}{v} h \bar{f} f=-\frac{m_{f}}{v} \sum_{i=1-3} \kappa_{if} h_{i} \bar{f} f, \\
\mathcal{L}_{h_{i} V V} &=\frac{1}{v} h\left(m_{Z}^{2} Z_{\mu} Z^{\mu}+2 m_{W}^{2} W_{\mu}^{+} W^{-\mu}\right)=\frac{1}{v} \sum_{i=1-3} \kappa_{iV} h_{i}\left(m_{Z}^{2} Z_{\mu} Z^{\mu}+2 m_{W}^{2} W_{\mu}^{+} W^{-\mu}\right),
\end{align}
where $m_f$ and $m_V$ are the same as those in the SM and Higgs coupling modifiers are
\begin{align}
\kappa_{if} = O_{1i},\quad \kappa_{iV} = O_{1i}.
\end{align}
In the SM limit, $\kappa_{1f}=\kappa_{1V}=1$ and $\kappa_{2,3f}=\kappa_{2,3V}=0$. We emphasize that the SM-like limit is also achieved in the degenerate scalar scenario, $m_{h_1}\simeq m_{h_2}\simeq m_{h_3}\equiv m_h$. In this case, signal strengths of the Higgs boson could be SM like. For example, the cross section of the process $gg\to h_i \to VV^*$ would be cast into the form 
\begin{align}
\sigma_{gg\to h_i \to VV^*} \simeq \sigma_{gg\to h}^{\text{SM}}
\left[
	\sum_i\frac{\kappa_{if}^2\kappa_{iV}^2}{\Gamma_{h_i}}
\right]\Gamma_{h\to VV^*}^{\text{SM}},
\end{align}
where $\sigma_{gg\to h}^{\text{SM}}$ is the production cross section of $gg\to h$ in the SM, and likewise $\Gamma_{h\to VV^*}^{\text{SM}}$ is the partial decay width of $h\to VV^*$ in the SM. $\Gamma_{h_i}$ denote the total decay widths of $h_i$ and a narrow width approximation is used since $\Gamma_{h_i}\ll m_{h_i}$. Here, the interference terms are neglected.\footnote{The interference terms could be important if $|m_{h_i}-m_{h_j}|\lesssim \Gamma_{h_i}+\Gamma_{h_j}$~\cite{Fuchs:2014ola,Das:2017tob} (for recent study, see Ref.~\cite{Sakurai:2022cki}). In our benchmark points, however, the smallest mass differences is 500 MeV and the sum of the total decay widths are at most $\Gamma_h^{\text{SM}}=4.1$ MeV~\cite{LHCHiggsCrossSectionWorkingGroup:2013rie}.} In this model, $\Gamma_{h_i}$ are well approximated by $\Gamma_{h_i} \simeq \kappa_i^2\Gamma_h^{\text{SM}}$, where $\kappa_{i}\equiv\kappa_{if} =\kappa_{iV}$. With this, the cross section has the form
\begin{align}
\sigma_{gg\to h_i \to VV^*} \simeq \sigma_{gg\to h}^{\text{SM}}\cdot \text{Br}_{h\to VV^*}^{\text{SM}}.
\end{align}
We note that the current experimental constraints on the Higgs total decay width are $\Gamma_h^{\text{exp}}<14.4$ MeV (ATLAS~\cite{ATLAS:2018jym}) and $\Gamma_h^{\text{exp}}=3.2^{+2.4}_{-1.7}$ MeV (CMS~\cite{CMS:2022ley}), which are not precise enough to constrain $\Gamma_{h_i}$ in our scenario at this moment.

In the degenerate scalar scenario, the mass matrix (\ref{Mij}) is simplified to
\begin{align}
(\mathcal{M}_S^2)_{ij}\simeq \sum_kO_{ik}O_{jk}m_h^2=\delta_{ij}m_h^2. 
\end{align}
One may consider the cases $|\delta_2|\ll1$ and $|d_2|\ll1$ to satisfy this condition. 
As discussed in Sec.~\ref{sec:ewpt}, however, the size of $\delta_2$ is closely related to the strong first-order EWPT.
Therefore, we have to take $v_S^{r, i}/v\ll 1$ while maintaining $\delta_2=\mathcal{O}(1)$ and $d_2=\mathcal{O}(1)$. 

Before closing this section, we make a comment on CPV in the CxSM.
Since $S$ couples to fermions and gauge bosons only through the mixing angles $\alpha_i$, a pseudoscalar coupling $h_i\bar{f}\gamma_5 f$ does not arise in this model. Therefore, even though the complex phases exist in the scalar potential and the singlet scalar VEV, they do not induce CPV in the matter sector in the SM, implying that further extension is needed to realize EWBG. One possible extension is adding new fermions that couple to $S$.
After integrating out the fermions, one may find the following higher dimensional operators contributing to the top Yukawa coupling
\begin{align}
\mathcal{L} = -y_t\bar{q}_L\tilde{H}\left(1+\frac{c_1}{\Lambda}S+\frac{c_2}{\Lambda^2}|S|^2+\frac{c_3}{\Lambda^2}S^2+\cdots\right)t_R+\text{H.c},\label{highDyt}
\end{align}
where $q_L$ is the left-handed doublet fermion, $\tilde{H}=i\tau^2H^*$ with $\tau^2$ being Pauli matrix, the coefficients $c_{1,2,3}$ are arbitrary complex parameters and $\Lambda$ is the scale of the integrated fermions. 
The imaginary parts of $c_1S$, $c_2$, and $c_3S^2$ give the pseudoscalar coupling, which may fuel EWBG.
The possibilities of EWBG using $\text{Im}(c_1)$ or $\text{Im}(c_2)$ in extensions of the SM with singlet scalar are studied in Refs.~\cite{Espinosa:2011eu,Cline:2012hg,Jiang:2015cwa,Cline:2021iff}. If all the coefficients $c_{1,2,3}$ happen to be real, the EWBG-related CPV should arise from $\text{Im}S$. This is the exactly the case we are interested in here. 
The possibility of spontaneous CPV is also studied in Ref.~\cite{Grzadkowski:2018nbc}, but the model setup is different from ours. 

%%%%%%%%%%%%%%%%%%%%%%%%%%%%%%%%%%%%%%%%%%%%%%%%%%%%%
%		            Electroweak phase transition
%%%%%%%%%%%%%%%%%%%%%%%%%%%%%%%%%%%%%%%%%%%%%%%%%%%%%
\section{Electroweak phase transition}\label{sec:ewpt}
We begin by discussing possbile patterns of phase transitions in this model. 
Under the assumption of $a_1^r\ne0$, one could find following phases:
\begin{align}
\text{EW}_1&: \big(v(T), v_S^r(T), v_S^i(T)\big); \\
\text{EW}_2&: \big(v(T), v_S^r(T), 0\big);\\
\text{S}'&: \big(0, \tilde{v}_S^r(T), \tilde{v}_S^i(T)\big); \\
\text{S}&: \big(0, \tilde{v}_S^r(T), 0\big),
\end{align}
where $T$ denotes temperature. The EW symmetry-broken phase comprises two cases $\text{EW}_1$ and $\text{EW}_2$ in which $v_S^i\neq0$ and $v_S^i=0$, respectively. Similarly, before the EW symmetry is broken, the vacua can be classified by the condition that spontaneous CPV exists or not, denoted as S$'$  and S, respectively.
Therefore, the possible thermal history in this model is as follows.
\begin{align}
\text{1 step PT}&:\quad \text{S}\to \text{EW}, \quad \text{S}'\to \text{EW}; \\
\text{2 step PT}&:\quad \text{S}\to \text{S}' \to \text{EW},
\end{align}
where EW denotes $\text{EW}_1$ or $\text{EW}_2$.
As seen from Eq.~(\ref{dvdchi}), $v_S^i$ would not be zero even at sufficiently high temperature if $a_1^i\neq0$, which we assume here to avoid the CP domain wall problem.
Therefore, the PT pattern in our case is $\text{S}': \big(0, \tilde{v}_S^r(T), \tilde{v}_S^i(T)\big) \to \text{EW}_1: \big(v(T), v_S^r(T), v_S^i(T)\big)$.

As far as EWBG is concerned, EWPT has to be strong first order such that~\cite{Arnold:1987mh,Bochkarev:1987wf,Funakubo:2009eg,Fuyuto:2014yia,Gan:2017mcv}
\begin{align}
\frac{v_{C}}{T_{C}} \gtrsim 1,\label{sph}
\end{align}
where $T_C$ is the critical temperature at which the effective potential has two degenerate minima, and $v_C$ is the doublet Higgs VEV at $T_C$. The lower value primarily depends on  saddle-point classical configuration called sphaleron~\cite{Manton:1983nd,Klinkhamer:1984di}. In the literature, the lower bound ``1" is commonly adopted as a rough criterion. 
As noted in Sec.~\ref{sec:model}, the current model must be extended to incorporate a new CPV. Even in such a case, the dynamics of EWPT and the lower value of Eq.~(\ref{sph}) would not be drastically changed if the new particles are fermions that are weakly coupled to $S$. Generally, the effects of the fermions appear at loop levels and thus subleading.

To study EWPT, we use a one-loop effective potential
\begin{align}
V_{1}\left(\varphi, \varphi_{S}^r, \varphi_{S}^i ; T\right)=\sum_{i} n_{i}\left[V_{\mathrm{CW}}\left(\bar{m}_{i}^{2}\right)+\frac{T^{4}}{2 \pi^{2}} I_{B, F}\left(\frac{\bar{m}_{i}^{2}}{T^{2}}\right)\right],
\end{align}
where $\varphi, \varphi_{S}^r$, and $\varphi_{S}^i$ are the constant classical background fields of $H$, $\text{Re}S$, and $\text{Im}S$, respectively. The indicies $i$ represent $h_{1,2,3}$, $W$, $Z$, $t$, and $b$. The degrees of freedom of each particle $n_i$ are $n_{h_{123}}=1$, $n_{G^0}=1$, $n_{G^\pm}=2$, $n_W=6$, $n_Z=3$, and $n_t=n_b=-12$. $V_{\text{CW}}$ is the $\overline{\text{MS}}$-defined effective potential at zero temperature, while $I_{\text{B,F}}$ are the nonzero temperature counterparts, which are, respectively, given by~\cite{Coleman:1973jx,Jackiw:1974cv,Dolan:1973qd}
\begin{align}
V_{\mathrm{CW}}\left(\bar{m}_{i}^{2}\right) &=\frac{\bar{m}_{i}^{4}}{64 \pi^{2}}\left(\ln \frac{\bar{m}_{i}^{2}}{\bar{\mu}^{2}}-c_{i}\right), \\
I_{B, F}\left(a^{2}\right) &=\int_{0}^{\infty} d x~x^{2} \ln \left(1 \mp e^{-\sqrt{x^{2}+a^{2}}}\right).
\end{align}
Here $\bar{m_i}$ denote field-dependent masses of each particle $i$, $c_i=3/2$ for scalars and fermions while $c_i=5/6$ for gauge bosons, $I_B$ with the upper sign is the thermal function for bosons while $I_F$ with the lower one is that for fermions.
$\bar{\mu}$ is a renormalization scale. In this work, we impose renormalization conditions such that the vacuum and scalar masses are not altered at the one-loop level, viz, on-shell-like scheme~\cite{Kirzhnits:1976ts}.\footnote{In this scheme, the NG contributions cause an infrared divergence, requiring a special treatment. In Ref.~\cite{Chiang:2018gsn}, it is found that their effects on EWPT are rather minor. We therefore exclude them in $V_{\text{eff}}$.}

To discuss the strong first-order EWPT qualitatively, we expand $I_{B,F}$ in powers of $\bar{m}^2/T^2$ and retain $\mathcal{O}(T^2)$ term (see Appendix~\ref{app:fm} for details).\footnote{We do not include the $T\varphi^3$-like term here since (i) $T\varphi^3$ is not the main source of the first-order EWPT in this model and (ii) CP-violating effect that we are interested in this paper arises from the tree-level potential rather than the thermal potential.
Therefore, retaining $\mathcal{O}(T^2)$ terms suffices for the qualitative study. 
}
Furthermore, we drop the zero-temperature one-loop corrections for simplicity. We call this simplified potential the high-temperature (HT) potential, which takes the form
\begin{align}
V^{\mathrm{HT}}\left(\varphi, \varphi_{S}^{r}, \varphi_{S}^{i} ; T\right)=&V_0(\varphi,\varphi_S^r,\varphi_S^i)+\frac{T^{2}}{2}\left[\Sigma_{H} \varphi^{2}+\Sigma_{S} \varphi_{S}^{r 2}+\Sigma_{S} \varphi_{S}^{i 2}\right] \nonumber\\ 
=& \frac{m^{2}}{4} \varphi^{2}+\frac{\lambda}{16} \varphi^{4}+\frac{\delta_{2}}{8} \varphi^{2}\left(\varphi_{S}^{r 2}+\varphi_{S}^{i 2}\right)+\frac{d_{2}}{16}\left(\varphi_{S}^{r 2}+\varphi_{S}^{i 2}\right)^{2}\nonumber \\ 
&+\sqrt{2}\left(a_{1}^{r} \varphi_{S}^{r}-a_{1}^{i} \varphi_{S}^{i}\right)+\frac{1}{4}b_{1}^{r}\left(\varphi_{S}^{r 2}-\varphi_{S}^{i 2}\right)+\frac{b_{2}}{4}\left(\varphi_{S}^{r 2}+\varphi_{S}^{i 2}\right)\nonumber \\ 
&+\frac{T^{2}}{2}\left[\Sigma_{H} \varphi^{2}+\Sigma_{S} \varphi_{S}^{r 2}+\Sigma_{S} \varphi_{S}^{i 2}\right],
\label{VHT}
\end{align}\\
where
\begin{align}
\Sigma_{H} &=\frac{\lambda}{8}+\frac{\delta_{2}}{24}+\frac{3}{16} g_{2}^{2}+\frac{1}{16} g_{1}^{2}+\frac{y_{t}^{2}}{4}+\frac{y_{b}^{2}}{4},\\
\Sigma_{S} &=\left(\delta_{2}+d_{2}\right) \frac{1}{12},
\end{align}
with $g_2$, $g_1$, $y_t$, and $y_b$ being the $\text{SU}(2)_L$, $\text{U}(1)_Y$ gauge couplings, top, and bottom Yukawa couplings, respectively. As analytically shown using the CPC HT potential~\cite{Cho:2021itv}, when additional singlet fields are present, the doublet-singlet mixing in the tree-level potential could contribute to the first-order EWPT significantly (for classification of first-order phase transitions, see, e.g., Ref.~\cite{chung2013125}). 
Here, we generalize it to the CPV case and see to what extent CPV can affect the strength of the first-order EWPT.

For the sake of convenience, the three scalar fields are reexpressed in terms of the polar coordinates as
\begin{align}
\varphi=z \cos \gamma, \quad \varphi_{S}^{r}=z \sin \gamma \cos \theta+\tilde{v}_{S}^{r}, \quad \varphi_{S}^{i}=z \sin \gamma \sin \theta+\tilde{v}_{S}^{i},
\end{align}
where the limit of $z=0$ describes to the S$'$ phase. 
A critical temperature ($T_C$) of the first-order EWPT is defined by a temperature at which the effective potential has two degenerated minima, such as 
\begin{align}
V(z_C,\gamma_C^{}, \theta_C^{}; T_C) = c_4z^2(z-z_C)^2,\quad z_C = \frac{c_3}{2c_4},
\end{align}
where the subscripts $C$ indicate that the quantities are evaluated at $T_C$, $c_3$ and $c_4$ are the coefficients of the cubic and quartic terms in $z$, i.e., $V^{\text{HT}}\ni -c_3z^3+c_4z^4$. Their explicit forms are, respectively, given by
\begin{align}
c_3 & = -\frac{s_{\gamma_C^{}} c_{\gamma_C^{}}^2}{4}\big(c_{\theta_C^{}}\tilde{v}_S^r+s_{\theta_C^{}}\tilde{v}_S^i\big)\big(\delta_2+d_2t_{\gamma_C^{}}^2\big), \\
c_4 & = \frac{c_{\gamma_C^{}}^4}{16}\big(\lambda+2\delta_2t_{\gamma_C^{}}^2 + d_2t_{\gamma_C^{}}^4\big),
\end{align}
where $s_{\theta_C}=\sin\theta_C$, $c_{\theta_C}=\cos\theta_C$ and 
\begin{align}
t_{\gamma_{C}^{}}&=\frac{v_{S C}^{r}-\tilde{v}_{S C}^{r}}{v_{C} c_{\theta_C^{}}}=\frac{v_{S C}^{i}-\tilde{v}_{S C}^{i}}{v_{C} s_{\theta_C^{}}}.
\label{tgamC}
\end{align}
In the case of $|t_{\gamma_C^{}}|\ll 1$, one could obtain simple expressions of $v_C$ and $T_C$ as
\begin{align}
v_C & \simeq \sqrt{\frac{2\delta_2}{\lambda}\Big(|\tilde{v}_{SC}|^2-\tilde{v}_{SC}^{i}(\tilde{v}_{SC}^{i}-t_{\theta_C^{}}\tilde{v}_{SC}^r)\Big)\left(1-\frac{v_{SC}^r}{\tilde{v}_{SC}^r}\right)}, \label{vc} \\
%& = \sqrt{\frac{2\delta_2}{\lambda}\Big(|\tilde{v}_{SC}|^2-\tilde{v}_{SC}^{r}(\tilde{v}_{SC}^{r}-t_{\theta_C^{}}\tilde{v}_{SC}^i)\Big)\left(1-\frac{v_{SC}^i}{\tilde{v}_{SC}^i}\right)} \\
T_{C} &\simeq \sqrt{\frac{1}{2 \Sigma_{H}}\left[-m^{2}-\frac{\delta_{2}}{2}\left|\tilde{v}_{S C}\right|^{2}\right]},\label{Tc}
\end{align}
where the first equality in Eq.~(\ref{tgamC}) is used to obtain $v_C$. The second equality in Eq.~(\ref{tgamC}) gives an equivalent expression of $v_C$, which is obtained by exchanging the indices $r$ and $i$ in Eq.~(\ref{vc}).
As is the CPC case studied in Ref.~\cite{Cho:2021itv}, the larger $\delta_2$ gives the more enhanced $v_C/T_C$, and moreover,
$T_C$ depends on $|\tilde{v}_{SC}|^2$ so that CPV does not matter. On the other hand, the degree to which $v_C$ is enhanced depends on $\tilde{v}_{SC}^{r}$, $\tilde{v}_{SC}^{i}$, and $\theta_C$. If $t_{\theta_C^{}}<0$, $v_C$ would get smaller compared to the CPC case. 

From the first equality in Eq.(\ref{del2}), one obtains
\begin{align}
\delta_{2} &= \frac{2}{v v_{S}^{r}} \sum_{i} O_{1 i} O_{2 i} m_{h_{i}}^{2}
 = \frac{2c_2}{v v_{S}^{r}}\Big[(m_{h_1}^2-m_{h_2}^2)s_1c_1c_3+s_2s_3(m_{h_3}^2-m_{h_1}^2c_1^2-m_{h_2}^2s_1^2) \Big].
\end{align}
In the degenerate scalar scenario, the numerator in $\delta_2$ gets small due to the orthogonality of the rotation matrix $O$, $\sum_iO_{1i}O_{2i}=0$. To compensate for it, $v_S^r$ has to be correspondingly small. One can get a similar expression from the second equality in Eq.(\ref{del2}) and find the necessity of the small $v_S^i$. As shown below, $v_S^r\lesssim 1$ GeV and $v_S^i\lesssim 1$ GeV are the typical sizes. The smallness of those VEVs controls the magnitudes and signs of $a_1^r$ and $a_1^i$ in order not to make $d_2$ too large, as seen from Eq.~(\ref{d2fromMM}). Overall, the necessary conditions for the strong first-order EWPT in CPV CxSM with the degenerate scalars are closely parallel with those in the CPC case. That said, the numerical values of $v_C$ could be distinctive in the presence of CPV. 
For instance, if we increase the CPV parameter $v_S^i$ while others are fixed, $\delta_2$ would get smaller, as discussed above, which could reduce $v_C$. We quantify this statement below.

Even though the HT potential makes it easy to extract the essence of EWPT, the dropped terms in the full potential $V_{\text{eff}}=V_0+V_1$ should be incorporated for quantitative studies. Furthermore, thermal resummation is needed to improve perturbative expansion at finite temperature. 
There exist two representative resummation methods called Carrington-Arnold-Espinosa~\cite{Carrington:1991hz,Arnold:1992rz} and Parwani~\cite{Parwani:1991gq} schemes.
In the former, only a zero Matsubara frequency mode is resummed, while all the modes are resummed in the latter.
Scheme dependence could be significant if the first-order EWPT is induced by a cubic term originating from $I_B$. In the CxSM, however, the structure of the tree-level potential plays a dominant role in realizing the strong first-order EWPT.   
It is verified in Ref.~\cite{Cho:2021itv} that $T_C$ and VEVs at this temperature in the two schemes are not different from each other significantly. Even though the current analysis includes CPV, the strong first-order EWPT is still controlled by the tree-level potential, and therefore we consider only the Parwani scheme for illustration.
A comment on our resummation scheme should be made here. 
In the original Parwani scheme, $\bar{m_i}^2$ are replaced with thermally corrected ones in all the places in $V_1$, which is the consequence of the reorganization of the perturbation theory in which mass parameters contain the temperature corrections at zeroth order.
On the other hand, we have renormalized $V_{\text{CW}}$ in the temperature-independent on-shell like scheme to maintain the tree-level relationships among the input model parameters. Since $V_{\text{CW}}$ is free from dangerous infrared divergences originating from the zero Matsubara mode, we perform the thermal resummation only in $I_B$, which is actually equivalent to a so-called daisy resummation shown in the pioneering work of Dolan and Jackiw~\cite{Dolan:1973qd}. 

In the above discussion, we define the transition temperature by $T_C$. 
To be more precise, however, the onset of the transition is bubble nucleation which occurs at a temperature lower than $T_C$. 
Now let us define the nucleation temperature $T_N$ and study to what extent this temperature deviates from $T_C$.

A bubble nucleation rate $\Gamma_N(T)$ per unit time and per unit volume may take the form~\cite{Linde:1981zj}
\begin{align}
\Gamma_{N}(T) \simeq T^{4}\left(\frac{S_3(T)}{2 \pi T}\right)^{3 / 2} e^{-S_3(T) / T},\label{GamN}
\end{align}
where $S_3(T)$ is three-dimensional bounce action describing an energy of a bubble that has a critical radius (critical bubble) at $T$. The nucleation temperature $T_N$ is defined by a relation 
\begin{align}
\frac{\Gamma_{N}\left(T_{N}\right)}{H^{3}\left(T_{N}\right)}
 =H\left(T_{N}\right) \simeq 1.66 \sqrt{g_{*}(T_{N})} \frac{T_{N}^{2}}{ m_{\mathrm{P}}},\label{TNdefinition}
\end{align}
where $H\left(T_{N}\right)$ is the Hubble parameter, $m_{\text{P}}\simeq 1.22\times 10^{-19}$ GeV, $g_{*}(T_{N})$ denotes the massless degrees of freedom at $T_N$, and we take $g_{*}(T_{N})=108.75$ in the numerical analysis. 
With $\Gamma_N(T)$ in Eq~(\ref{GamN}), Eq.(\ref{TNdefinition}) is recast into the form
\begin{align}
\frac{S_3\left(T_{N}\right)}{T_{N}}-\frac{3}{2} \ln \left(\frac{S_3\left(T_{N}\right)}{T_{N}}\right)=143.4-2 \ln\left(\frac{g_{*}\left(T_{N}\right)}{100}\right)-4 \ln \left(\frac{T_{N}}{100\ \mathrm{GeV}}\right),
\end{align}
which implies that $S_3(T_N)/T_N\lesssim140$ would be required for the occurrence of EWPT.
There exists a tendency that the stronger first-order EWPT induces the larger $S_3/T$, and beyond certain strength of the transition, the condition $S_3/T\lesssim 140$ would not be satisfied even at sufficiently low temperature. We rule out such a case in our analysis.\footnote{EWPT may be triggered by QCD phase transition~\cite{Witten:1980ez} (for recent studies, see, e.g., Refs~\cite{Iso:2017uuu,Arunasalam:2017ajm,Bodeker:2021mcj,Kawana:2022fum}). Besides this possibility, it is pointed out in Ref.~\cite{Blasi:2022woz} that domain walls, if exist, can catalyze the bubble nucleation.
}

We obtain the critical bubble using the following energy functional
\begin{align}
S_3(T)=\int d^{3} x\left[\left(\partial_{i} H\right)^{\dagger} \partial_{i} H+\partial_{i} S^{*} \partial_{i} S+V_{\text{eff}}(H, S ; T)\right],
\label{energyfunctional}
\end{align}
where the classical background fields are parameterized as
\begin{align}
\langle H(x)\rangle=\frac{1}{\sqrt{2}}\left(\begin{array}{c}
0 \\
\rho(x)
\end{array}\right), \quad\langle S(x)\rangle=\frac{1}{\sqrt{2}}\left(\rho_{S}^{r}(x)+i \rho_{S}^{i}(x)\right).
\end{align}
Since a spherically symmetric configuration is expected to give the least energy, the scalar fields depend only on the radial coordinate $r=\sqrt{x^2+y^2+z^2}$. In this case, the energy functional (\ref{energyfunctional}) takes the simplified form
\begin{align}
S_3(T)=4 \pi \int_{0}^{\infty} d r\ r^{2}\left[\frac{1}{2}\left(\frac{d \rho}{d r}\right)^{2}+\frac{1}{2}\left(\frac{d \rho_{S}^{r}}{d r}\right)^{2}+\frac{1}{2}\left(\frac{d \rho_{S}^{i}}{d r}\right)^{2}+\bar{V}_{\text{eff}}\left(\rho, \rho_{S}^{r}, \rho_{S}^{i} ; T\right)\right],\label{S3}
\end{align}
where we normalize the potential as 
\begin{align}
\bar{V}_{\text{eff}}\left(\rho, \rho_{S}^{r}, \rho_{S}^{i} ; T\right)=V_{\text{eff}}\left(\rho, \rho_{S}^{r}, \rho_{S}^{i} ; T\right)-V_{\text{eff}}\left(0, \tilde{v}_{S}^{r}, \tilde{v}_{S}^{i} ; T\right).
\end{align}
From this energy functional, one obtains the equations of motion (EOMs) as
\begin{align}
\frac{d^{2} \rho}{d r^{2}}+\frac{2}{r} \frac{d \rho}{d r}-\frac{\partial \bar{V}}{\partial \rho}=0, \\
\frac{d^{2} \rho_{S}^{r}}{d r^{2}}+\frac{2}{r} \frac{d \rho_{S}^{r}}{d r}-\frac{\partial \bar{V}}{\partial \rho_{S}^{r}}=0,\label{EOM_rhoSr} \\
\frac{d^{2} \rho_{S}^{i}}{d r^{2}}+\frac{2}{r} \frac{d \rho_{S}^{i}}{d r}-\frac{\partial \bar{V}}{\partial \rho_{S}^{i}}=0.\label{EOM_rhoSi}
\end{align}
For the critical bubble, the boundary conditions are
\begin{align}
&\lim _{r \rightarrow \infty} \rho(r)=0, \quad \lim _{r \rightarrow \infty} \rho_{S}^{r}(r)=\tilde{v}_{S}^{r}, \quad \lim _{r \rightarrow \infty} \rho_{S}^{i}(r)=\tilde{v}_{S}^{i}, \label{bc1} \\
&\left.\frac{d \rho(r)}{d r}\right|_{r=0}=0,\left.\quad \frac{d \rho_{S}^{r}(r)}{d r}\right|_{r=0}=0,\left.\quad \frac{d \rho_{S}^{i}(r)}{d r}\right|_{r=0}=0.\label{bc2}
\end{align}
By the first boundary conditions (\ref{bc1}), $\bar{V}$ approaches to zero in the symmetric phase ($r\to \infty$) and $S_3$ remains finite. The second boundary conditions (\ref{bc2}) are needed to avoid the singularities of EOMs at the broken phase ($r=0$). 
We note in passing that when dividing $S_3$ into the kinetic and potential parts, $S_3=T+U$, the classical solution satisfies the relation $T=-3U$ by the Derrick theorem~\cite{Derrick:1964ww}. This gives a useful cross-check for the correctness of numerical solutions. 

Now we present our numerical results. To analyze EWPT, we use a public code \texttt{CosmoTransitions}~\cite{Wainwright:2011kj}. In finding $T_N$, we simply employ the condition $S_3(T_N)/T_N=140$.
To see the impacts of CPV on EWPT and make a comparison with the CPC case studied in Ref.~\cite{Cho:2021itv}, we closely follow the parameter choices in Ref.~\cite{Cho:2021itv}, where $v_S^r =0.6$ GeV, $m_{h_2}=124.0$ GeV, and $\alpha_1=\pi/4$ are chosen. For other input parameters, we set $m_{h_3}=124.5$ GeV and $\alpha_2=0.0$ radians for illustration.

%------------------------------------------------------------------------------------------------------
\begin{figure}[t]
\center
\includegraphics[width=8cm]{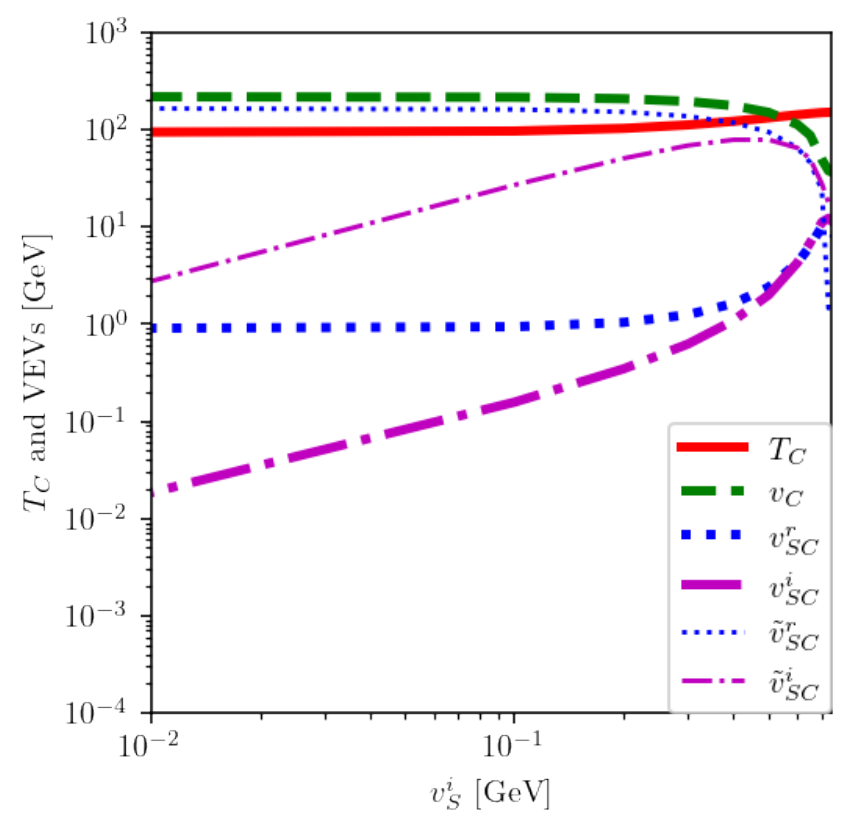}
\caption{Shown here are $T_C$ (solid, red), $v_C$ (dashed, green), $v_{SC}^r$ (thick-dotted, blue), $\tilde{v}_{SC}^r$ (thin-dotted, blue), $v_{SC}^i$ (thick-dot-dashed, magenta), and $\tilde{v}_{SC}^i$ (thin-dot-dashed, magenta) as functions of $v_S^i$. The other input parameters are listed in Table~\ref{tab:BP}.
}
\label{fig:EWPT_vsi}
\end{figure}
%------------------------------------------------------------------------------------------------------

Fig.~\ref{fig:EWPT_vsi} shows VEVs in the symmetric and broken phases at $T_C$ as functions of $v_S^i$, where line and color schemes represent  $T_C$ (solid, red), $v_C$ (dashed, green), $v_{SC}^r$ (thick-dotted, blue), $\tilde{v}_{SC}^r$ (thin-dotted, blue), $v_{SC}^i$ (thick-dot-dashed, magenta), and $\tilde{v}_{SC}^i$ (thin-dot-dashed, magenta), respectively. One can see that $v_{SC}^i$ and $\tilde{v}_{SC}^i$ are very sensitive to $v_S^i$. The former is the monotonically increasing function, while the latter has the similar behavior but turns around $v_S^i\simeq 0.5$ and drops.
This result implies that sizable CPV at $T_C$ requires the sizable CPV at $T=0$ GeV. In other words, the so-called \textit{transitional CPV}~\cite{Funakubo:1995kw,Funakubo:1999ws,Huber:1999sa,Huber:2000mg,Huber:2001xf}, in which CPV can be sizable at $T_C$ but suppressed or even zero at $T=0$ GeV, does not occur in the current case. 
Actually, we find that $\theta_S(T)=\tan^{-1}[v_S^i(T)/v_S^r(T)]$ is temperature independent, which may be understood as follows. 
The phase-dependent part of the HT potential is 
\begin{align}
V^{\text{HT}}(\vartheta_S) &= \sqrt{2}(a_1^r\varphi_S^r-a_1^i\varphi_S^i) +\frac{1}{4}b_{1}^{r}\left(\varphi_{S}^{r 2}-\varphi_{S}^{i 2}\right) \nonumber\\
&= \sqrt{2}\varphi_S(a_1^r\cos\vartheta_S-a_1^i\sin\vartheta_S)
+\frac{1}{4}b_{1}^{r}\varphi_S^2\left(\cos^2\vartheta_S-\sin^2\vartheta_S\right), \label{VHTth}
\end{align}
where $\varphi_S^r=\varphi_S\cos\vartheta_S$ and $\varphi_S^i=\varphi_S\sin\vartheta_S$.
Even though $\varphi_S$ evolves with temperature, $\vartheta_S$ does not since the temperature-dependent structure is $T^2\Sigma_S(\varphi_S^{r2}+\varphi_S^{i2})=T^2\Sigma_S\varphi_S^2$, and thus $\langle \vartheta_S(T)\rangle=\theta_S(T)$ stays on its zero temperature value $\theta_S(T=0)=\tan^{-1}(v_S^i/v_S^r)$.
This argument can also apply to the case using the full effective potential. 

The $v_S^i$ dependence on the other VEVs and $T_C$ are rather mild for $v_S^i\lesssim 0.3$ GeV. This may be due to the fact that the leading CPV effect, which enters through the term $-\sqrt{2}a_1^i\varphi_S^i$ in the tree-level potential (\ref{VHTth}),\footnote{$b_1^r$ is negligibly small in the chosen parameter space. See Table~\ref{tab:BP}.} has less effect compared to the term $\sqrt{2}a_1^r\varphi_S^r$ for $v_S^i\lesssim 0.3$ GeV. Beyond it, however, the CPV piece becomes comparable to the CPC one and its effect starts to get pronounced. It is found that $v_C/T_C$ becomes weaker as $v_S^i$ increases, leading to $v_C/T_C<1$ for $v_S^i\gtrsim 0.5$. The reduction of $v_C/T_C$ is the consequence of the diminution of $\delta_2\propto 1/v_S^i$, which can be understood by the analytic formulas Eqs.(\ref{vc}) and (\ref{Tc}).

As shown below, the bubble nucleation does not occur for $v_S^i<0.3$ GeV; thus, we exclusively focus on the cases $v_S^i = 0.3$, 0.4, and 0.5 GeV while keeping others fixed, which are referred to as BP1, BP2, and BP3, respectively. 
The input and output parameters in those benchmark points are summarized in Table~\ref{tab:BP}.
Note that Eq.~(\ref{b1r}) with $b_1^r\simeq 10^{-12}~\text{GeV}^2$ yields $a_1^i \simeq -a_1^rv_S^i/v_S^r=-a_1^r\tan\theta_S$.
From Eq.~(\ref{alp3}), it is found that $\alpha_3=0.464, 0.588, 0.695$ radians in BP1, BP2, and BP3, respectively. 
In all the BPs, $\kappa_{1}=0.711$, $\kappa_2=-0.711$, and $\kappa_3=0.0$, which implies that $h_3$ does not couple to SM particles and only couples to $h_2$.

$T_C$ and the corresponding VEVs in the three BPs are given in Table~\ref{tab:atTc}. We also refer to the results in the CPC cases quoted from our previous work~\cite{Cho:2021itv}. Note that $\chi$ plays a role as DM, and the viable mass windows are only the two locations shown here. 
Even though some numerical differences are observed in  BP1 and CPC cases, they give a similar EWPT since CPV is not much effective in BP1. 

%----------------------------------------------------------------------------------------------------------------------------------
\begin{table}[t]
\center
\begin{tabular}{|c|c|c|c|c|c|c|c|c|}
\hline
Inputs & $v$ [GeV] & $v_S^r$ [GeV] & $v_S^i$ [GeV] & $m_{h_1}$ [GeV] & $m_{h_2}$ [GeV] & $m_{h_3}$ [GeV] & $\alpha_1$ [rad] & $\alpha_2$ [rad] \\ \hline
BP1 & 246.22 & 0.6 & 0.3 & 125.0 & 124.0 & 124.5 & $\pi/4$ & 0.0  \\ \hline
BP2 & 246.22 & 0.6 & 0.4 & 125.0 & 124.0 & 124.5 & $\pi/4$ & 0.0  \\ \hline
BP3 & 246.22 & 0.6 & 0.5 & 125.0 & 124.0 & 124.5 & $\pi/4$ & 0.0  \\ \hline\hline
Outputs & $m^2$ & $b_2$ [GeV$^2$] & $b_1^r$ [GeV$^2$] & $\lambda$ & $\delta_2$ & $d_2$ & $a_1^r$ [GeV$^3$] & $a_1^i$ [GeV$^3$]  \\ \hline
BP1 & $-(124.5)^2$ & $-(121.2)^2$ & $-7.717\times 10^{-12}$& 0.511 &1.51 & 1.111 & $-(18.735)^3$ & $(14.870)^3$ \\ \hline
BP2 & $-(124.5)^2$ & $-(107.3)^2$ & $5.145\times 10^{-12}$& 0.511 & 1.40 & 0.962 & $-(18.735)^3$ & $(16.367)^3$  \\ \hline
BP3 & $-(124.5)^2$ & $-(90.82)^2$ & $0.0000$& 0.511& 1.29 & 0.820 &$-(18.735)^3$ & $(17.630)^3$ \\ \hline
\end{tabular}
\caption{Input and output parameters in the three BPs. $\alpha_3=0.464, 0.588, 0.695$ radians in BP1, BP2, and BP3, respectively. The Higgs coupling modifiers are common to all the BPs, i.e., $\kappa_{1}=0.711$, $\kappa_2=-0.711$, and $\kappa_3=0.0$.}
\label{tab:BP}
\end{table}
%----------------------------------------------------------------------------------------------------------------------------------

%----------------------------------------------------------------------------------------------------------------------------------
\begin{table}[t]
\center
\begin{tabular}{|c|c|c|c|c|c|}
\hline
 & \multicolumn{3}{c|}{CPV} & \multicolumn{2}{c|}{CPC} \\ \hline
 & $v_S^i=0.3$ GeV & $v_S^i=0.4$ GeV & $v_S^i=0.5$ GeV & $m_\chi=62.5$ GeV & $m_\chi=2$ TeV \\ \hline
$v_C/T_C$ & $\frac{196.1}{112.3}=1.7$ & $\frac{177.2}{122.5}=1.4$ & $\frac{150.9}{132.8}=1.1$ & $\frac{200.1}{106.1}=1.9$  & $\frac{205.3}{108.7}=1.9$\\ 
$v_{SC}^r$ [GeV] & 1.249 & 1.634 & 2.403 & 1.250  & 1.171\\
$v_{SC}^i$ [GeV]  & 0.624 &1.089  & 2.003 & ---  & --- \\  
$\tilde{v}_{SC}^r$ [GeV]  & 137.9 & 118.5 & 94.82 & 144.2 & 146.2 \\
$\tilde{v}_{SC}^i$ [GeV]  &68.97  & 79.01 & 79.01 & ---  &---  \\  
\hline
\end{tabular}
\caption{VEVs at critical temperature $T_C$ in the three BPs and two CPC cases.  For details of the CPC cases, see Ref.~\cite{Cho:2021itv}.}
\label{tab:atTc}
\end{table}
%----------------------------------------------------------------------------------------------------------------------------------

Now we show the numerical results of the bubble nucleation. 
$T_N$ and the corresponding VEVs in the three BPs and CPC cases are summarized in Table~\ref{tab:atTn}.
The stronger first-order EWPT leads to the larger supercooling, which may be characterized by the quantity $\Delta = (T_C-T_N)/T_C$. 
One finds that $\Delta = 40.5\%$ for BP1, $\Delta = 16.7\%$ for BP2, and $\Delta = 7.3\%$ for BP3. Too large supercooling prevents EWPT from developing since $S_3/T>140$ at sufficiently low temperature, which occurs in the region $v_S^i<0.3$ GeV, as stated above.
For the CPC cases, it is found that $\Delta = 46.0\%$ for $m_\chi=62.5$ GeV, and $\Delta = 46.7\%$ for $m_\chi=2$ TeV. 
Despite the larger supercooling compared to BP1, we still have bubble nucleations in those cases. In passing, it turns out that the HT cases in Ref.~\cite{Cho:2021itv}, which give the even stronger first-order EWPT, do not have the bubble nucleation.

%----------------------------------------------------------------------------------------------------------------------------------
\begin{table}[t]
\center
\begin{tabular}{|c|c|c|c|c|c|}
\hline
 & \multicolumn{3}{c|}{CPV} & \multicolumn{2}{c|}{CPC} \\ \hline
 & $v_S^i=0.3$ GeV & $v_S^i=0.4$ GeV & $v_S^i=0.5$ GeV & $m_\chi=62.5$ GeV & $m_\chi=2$ TeV \\ \hline
$v_N/T_N$ & $\frac{239.0}{66.85}=3.6$ & $\frac{211.7}{102.0}=2.1$ & $\frac{177.2}{123.1}=1.4$ & $\frac{241.8}{57.20}=4.2$ & $\frac{242.4}{57.99}=4.2$\\ 
$v_{SN}^r$ [GeV]  & 0.657 & 0.921 &1.446  & 0.636 & 0.634 \\
$v_{SN}^i$ [GeV]  & 0.328  & 0.614 & 1.205 & --- & ---\\
$\tilde{v}_{SN}^r$ [GeV]  & 143.7 & 122.3 & 97.26 & 150.1 & 150.2\\
$\tilde{v}_{SN}^i$ [GeV]  &71.83  & 81.55 & 81.05 & --- & ---\\
$\Delta$ & 40.5\%  & 16.7\% & 7.3\% & 46.0\% & 46.7\% \\
\hline
\end{tabular}
\caption{VEVs at nucleation temperature $T_N$ in the three BPs and two CPC cases. $\Delta=(T_C-T_N)/T_C$, which characterizes the degrees of supercooling.}
\label{tab:atTn}
\end{table}
%----------------------------------------------------------------------------------------------------------------------------------

\begin{figure}
\center
\includegraphics[width=5.4cm]{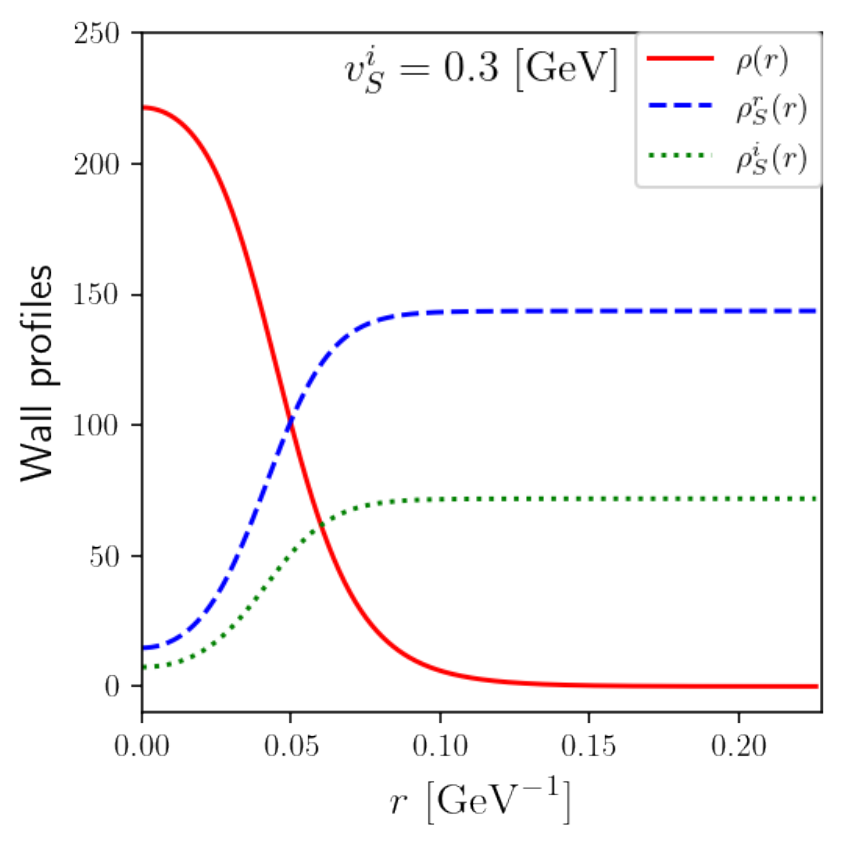}
\includegraphics[width=5.4cm]{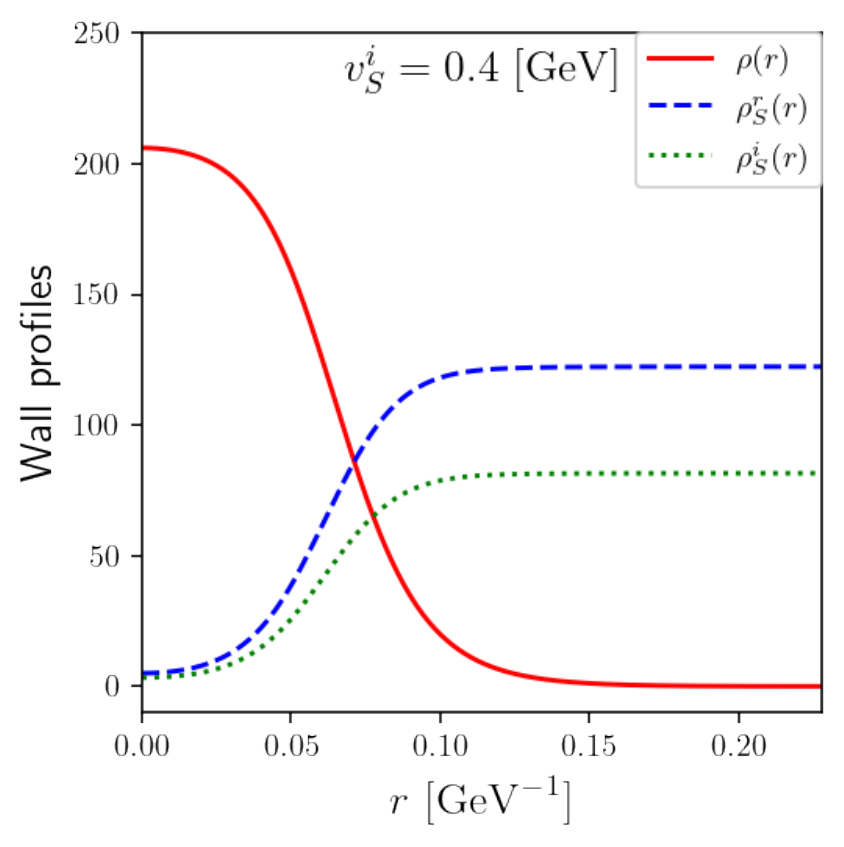}
\includegraphics[width=5.4cm]{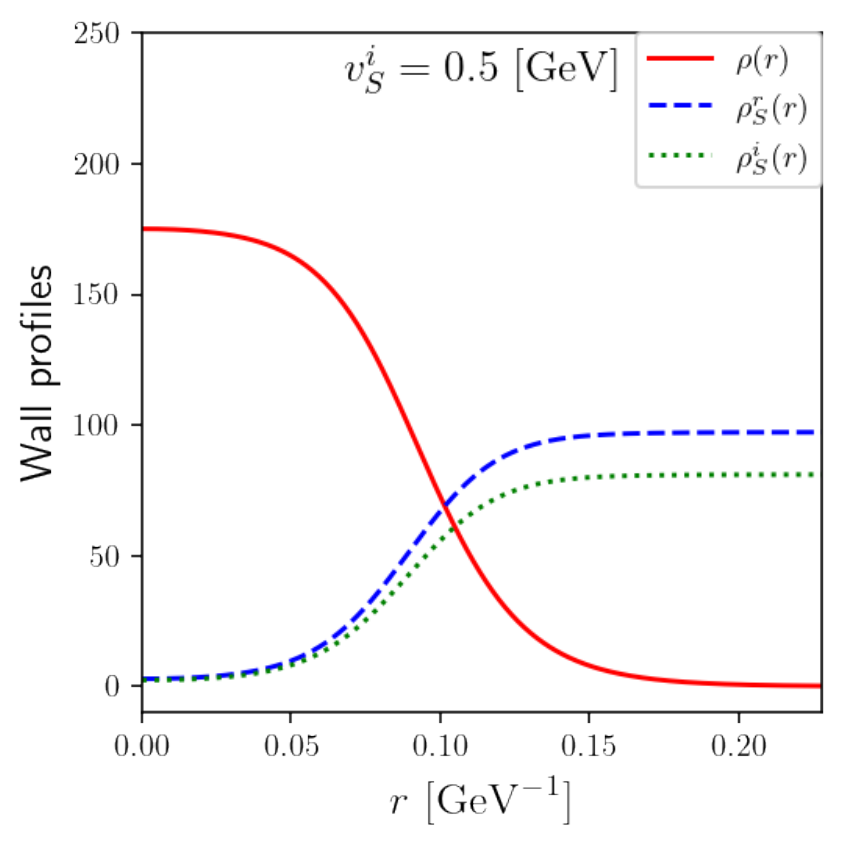}
\caption{$\rho(r)$ (solid, red), $\rho_S^r(r)$ (dashed, blue), and $\rho_S^i(r)$ (dotted, green) in BP1 (left), BP2 (middle), BP3 (right).}
\label{fig:profile}
\end{figure}

\begin{figure}
\center
\includegraphics[width=8cm]{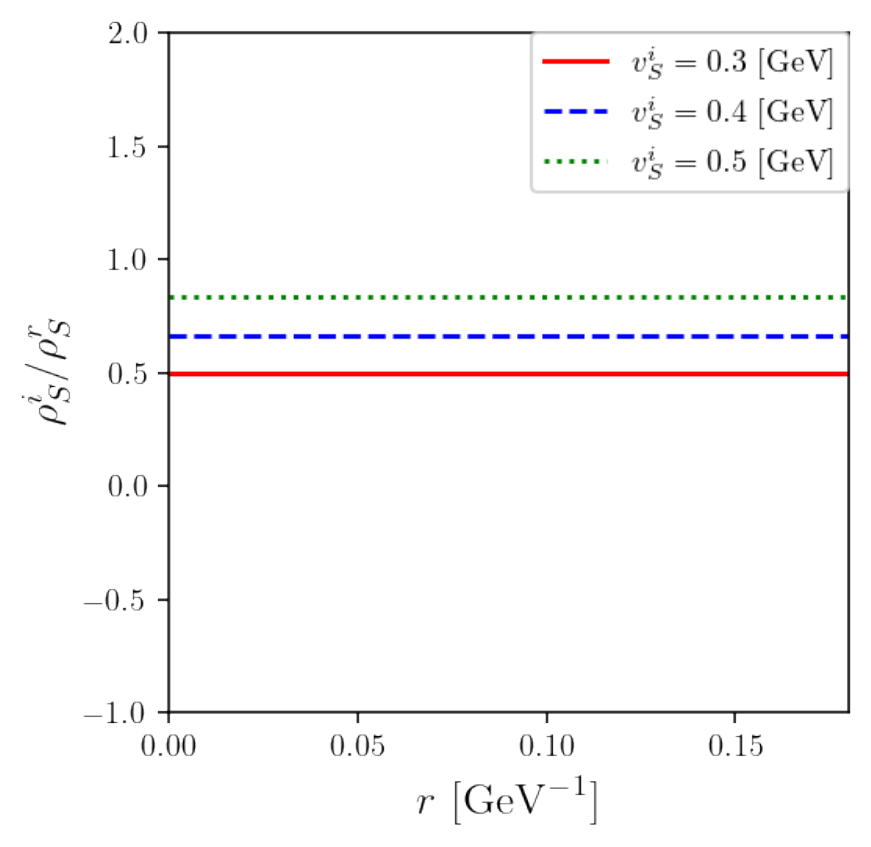}
\caption{$\rho_S^i/\rho_S^r$ as a function of $r$ in BP1 (solid, red), BP2 (dashed, blue), and PB3 (dotted, green).}
\label{fig:ratio}
\end{figure}

In the context of EWBG, the bubble wall profile is one of the most important key parameters in generating BAU.
Fig.~\ref{fig:profile} shows $\rho(r)$ (solid, red), $\rho_S^r(r)$ (dashed, blue), and $\rho_S^i(r)$ (dotted, green) in BP1 (left panel), BP2 (middle panel), and BP3 (right panel), respectively. The left endpoints correspond to the VEVs in the broken phase, while the right endpoints approach those in the symmetric phase. All the profiles have hyperbolic tangent shapes. Besides the endpoints, one crucial difference among the BPs is the thicknesses of the walls ($L_w$).
One can see that $L_w$ gets smaller as the strength of the first-order EWPT becomes larger. 
Commonly, it is called a thick wall regime if $L_w>1/T$ and otherwise a thin wall regime.
In BP1, $L_w\simeq 0.1~\text{GeV}^{-1}$, which is greater than $1/T_N\simeq 0.01~\text{GeV}^{-1}$, and hence thick wall regime. 
In Fig.~\ref{fig:ratio}, we also show the $r$ dependence on the phase of the singlet scalar profile.
It is found that $\rho_S^i/\rho_S^r$ in all the BPs are flat, which would be due to the same reason for the temperature independence of $\theta_S$ discussed around Eq.~(\ref{VHTth}), namely, $\theta_S$ is not dynamical in our model.

%------------------------------------------------------------------------------------------------------
%------------------------------------------------------------------------------------------------------

%------------------------------------------------------------------------------------------------------
%------------------------------------------------------------------------------------------------------

In closing, we briefly discuss CPV relevant to EWBG. As noted in Sec.~\ref{sec:model}, our model has to be extended to propagate CPV in the scalar sector to the SM fermion sector. For illustrative purpose, we consider the dimension 5 operator in Eq.~(\ref{highDyt}).
The top mass during EWPT is cast into the form
\begin{align}
m_t(r) = \frac{y_t\rho(r)}{\sqrt{2}}\left(1+\frac{c_1}{\sqrt{2}\Lambda}\big(\rho_S^r(r)+i\rho_S^i(r)\big)\right)\equiv |m_t(r)|e^{i\theta_t (r)},
\end{align}
where the phase $\theta_t(r)$ is expressed as
\begin{align}
\theta_t(r) & = \tan^{-1}\left(\frac{\rho_S^i(r)}{\sqrt{2}\Lambda/c_1+\rho_S^r(r)}\right).
\end{align}
The CPV source term for BAU can arise from the derivatives of $\theta_t(r)$ with respect to $r$. 
It is interesting to investigate to what extent BAU can be sourced. However, the detailed analysis of EWBG is beyond the scope of this paper, and we defer it to future studies.

%%%%%%%%%%%%%%%%%%%%%%%%%%%%%%%%%%%%%%%%%%%%%%%%%%%%%
%		            
%%%%%%%%%%%%%%%%%%%%%%%%%%%%%%%%%%%%%%%%%%%%%%%%%%%%%
\section{Gravitational waves}\label{sec:gw}
One of the important consequences of the strong first-order EWPT is GWs arising from bubble and plasma dynamics.
The amplitudes and frequencies of GWs would be modulated according to the amount of latent heat and/or duration of the phase transition. Those quantities may be quantified by the so-called $\alpha$ and $\beta$ parameters~\cite{Grojean:2006bp,Caprini:2015zlo}
\begin{align}
\alpha \equiv \frac{\epsilon\left(T_{*}\right)}{\rho_{\text {rad }}\left(T_{*}\right)},\quad \left.\beta \equiv H_{*} T_{*} \frac{d}{d T}\left(\frac{S_{3}(T)}{T}\right)\right|_{T=T_{*}},
\end{align}
with
\begin{align}
\epsilon(T)=\Delta V_{\mathrm{eff}}-T \frac{\partial \Delta V_{\mathrm{eff}}}{\partial T},\quad \rho_{\mathrm{rad}}(T)=\frac{\pi^{2}}{30} g_{*}(T) T^{4},
\end{align}
where $\Delta V_{\mathrm{eff}}=V_{\mathrm{eff}}\left(0, \tilde{v}_{S}^{r}(T), \tilde{v}_{S}^{i}(T) ; T\right)-V_{\mathrm{eff}}\left(v(T), v_{S}^{r}(T), v_{S}^{i}(T) ; T\right)$ and $H_{*}=H\left(T_{*}\right)$. 
$T_*$ is the temperature at which GWs are produced. Without significant reheating, $T_*$ would not much differ from $T_N$, and $T_{*}=T_{N}$ is taken in this study.

Sources of GWs arise from bubble collisions~\cite{PhysRevD.45.4514,PhysRevLett.69.2026,Kosowsky:1992vn,Kamionkowski:1993fg,Caprini:2007xq,Huber:2008hg}, sound waves~\cite{Hindmarsh:2013xza,Giblin:2013kea,Giblin:2014qia,Hindmarsh:2015qta} and turbulence induced by percolation~\cite{Caprini:2006jb,Kahniashvili:2008pf,Kahniashvili:2008pe,Kahniashvili:2009mf,Caprini:2009yp,Binetruy:2012ze}.
The GW spectrum we observe is the sum of them, i.e., 
\begin{align}
\Omega_{\mathrm{GW}}(f) h^{2}=\Omega_{\mathrm{col}}(f) h^{2}+\Omega_{\mathrm{sw}} (f)h^{2}+\Omega_{\mathrm{turb}}(f) h^{2},
\end{align}
where $f$ is the frequency of GW. We estimate $\Omega_{\mathrm{GW}}$ using the following equations~\cite{Grojean:2006bp,Caprini:2015zlo}

\begin{align}
\Omega_{\mathrm{col}}h^{2}=1.67 \times 10^{-5}\left(\frac{\beta}{H_{*}}\right)^{-2}\left(\frac{\kappa_{\mathrm{col}} \alpha}{1+\alpha}\right)^{2}\left(\frac{100}{g_{*}}\right)^{1 / 3}\left(\frac{0.11 v_{w}^{3}}{0.42+v_{w}^{2}}\right) \frac{3.8\left(f / f_{\mathrm{col}}\right)^{2.8}}{1+2.8\left(f / f_{\mathrm{col}}\right)^{3.8}},\label{Oh2col} \\
\Omega_{\mathrm{sw}} h^{2}=2.65 \times 10^{-6}\left(\frac{\beta}{H_{*}}\right)^{-1}\left(\frac{\kappa_{v} \alpha}{1+\alpha}\right)^{2}\left(\frac{100}{g_{*}}\right)^{1 / 3} v_{w}\left(\frac{f}{f_{\mathrm{sw}}}\right)^{3}\left(\frac{7}{4+3\left(f / f_{\mathrm{sw}}\right)^{2}}\right)^{7 / 2},\label{Oh2sw}\\
\Omega_{\mathrm{turb}} h^{2}=3.35 \times 10^{-4}\left(\frac{\beta}{H_{*}}\right)^{-1}\left(\frac{\kappa_{\mathrm{turb}} \alpha}{1+\alpha}\right)^{\frac{3}{2}}\left(\frac{100}{g_{*}}\right)^{1 / 3} v_{w} \frac{\left(f / f_{\text {turb }}\right)^{3}}{\left[1+\left(f / f_{\mathrm{turb}}\right)\right]^{\frac{11}{3}}\left(1+8 \pi f / h_{*}\right)},\label{Oh2turb}
\end{align}
where $v_w$ denotes the bubble wall velocity and $h_*$ is
\begin{align}
h_{*}&=1.65 \times 10^{-5}\left(\frac{T_{*}}{100~\mathrm{GeV}}\right)\left(\frac{g_{*}}{100}\right)^{1 / 6} \mathrm{~Hz}.
\end{align}
From Eqs.~(\ref{Oh2col}), (\ref{Oh2sw}), and (\ref{Oh2turb}), one can see that $\Omega_{\text{GW}} h^2$ becomes more enhanced as $\alpha$ increases and/or $\beta$ decreases. Since the larger supercooling corresponds to the larger $\alpha$, we can expect that $\Omega_{\text{GW}}$ in BP1 or CPC cases becomes the highest among others. 
We also note that the powers of $\beta/H_*$ in $\Omega_{\text{sw}}h^2$ and $\Omega_{\text{turb}}h^2$ are higher than that in $\Omega_{\text{col}}h^2$, which is attributed to the fact that the former two have the longer duration of the sources. As numerically shown below, the sound wave contribution is dominant in the vicinity of the peak frequency. 

For $v_w\simeq 1$, it is found that
\begin{align}
\kappa_{\mathrm{col}} &\simeq \frac{1}{1+0.715 \alpha}\left(0.715 \alpha+\frac{4}{27} \sqrt{\frac{3 \alpha}{2}}\right), \\
\kappa_{v} &\simeq \frac{\alpha}{0.73+0.083 \sqrt{\alpha}+\alpha}, \\
\kappa_{\mathrm{turb}} & \simeq (0.05-0.1) \kappa_{v}.
\end{align}
In our numerical analysis, we take $v_w=0.95$ and $\kappa_{\text{turb}}=0.1\kappa_v$ for illustration. The other choices, e.g., $\kappa_{\text{turb}}=0.05\kappa_v$,  do not change our conclusion.

The peak frequencies of the three sources are, respectively, given by
\begin{align}
f_{\mathrm{col}} &=16.5 \times 10^{-6} \left(\frac{\beta}{H_{*}}\right)\left(\frac{0.62}{1.8-0.1 v_{w}+v_{w}^{2}}\right)\left(\frac{T_{*}}{100~\mathrm{GeV}}\right)\left(\frac{g_{*}}{100}\right)^{1 / 6}\  \mathrm{~Hz}, \\
f_{\mathrm{sw}} &=1.9 \times 10^{-5}\frac{1}{v_{w}}\left(\frac{\beta}{H_{*}}\right)\left(\frac{T_{*}}{100~\mathrm{GeV}}\right)\left(\frac{g_{*}}{100}\right)^{1 / 6}\  \mathrm{~Hz}, \\
f_{\text {turb }} &=2.7 \times 10^{-5} v_{w}^{-1} \left(\frac{\beta}{H_{*}}\right)\left(\frac{T_{*}}{100~\mathrm{GeV}}\right)\left(\frac{g_{*}}{100}\right)^{1 / 6} \mathrm{~Hz}.
\end{align}
Even though $\Omega_{\text{sw}}$ is the biggest contribution in $\Omega_{\text{GW}}$ around the peak frequency, 
the other two contributions would come into play at higher frequencies since $\Omega_{\text{sw}}\propto f^{-4}$ while $\Omega_{\text{col}}\propto f^{-1}$ and $\Omega_{\text{turb}}\propto f^{-5/3}$~\cite{Caprini:2015zlo}. 
We, therefore, include all the contributions in our numerical analysis.

\begin{figure}
\center
\includegraphics[width=8cm]{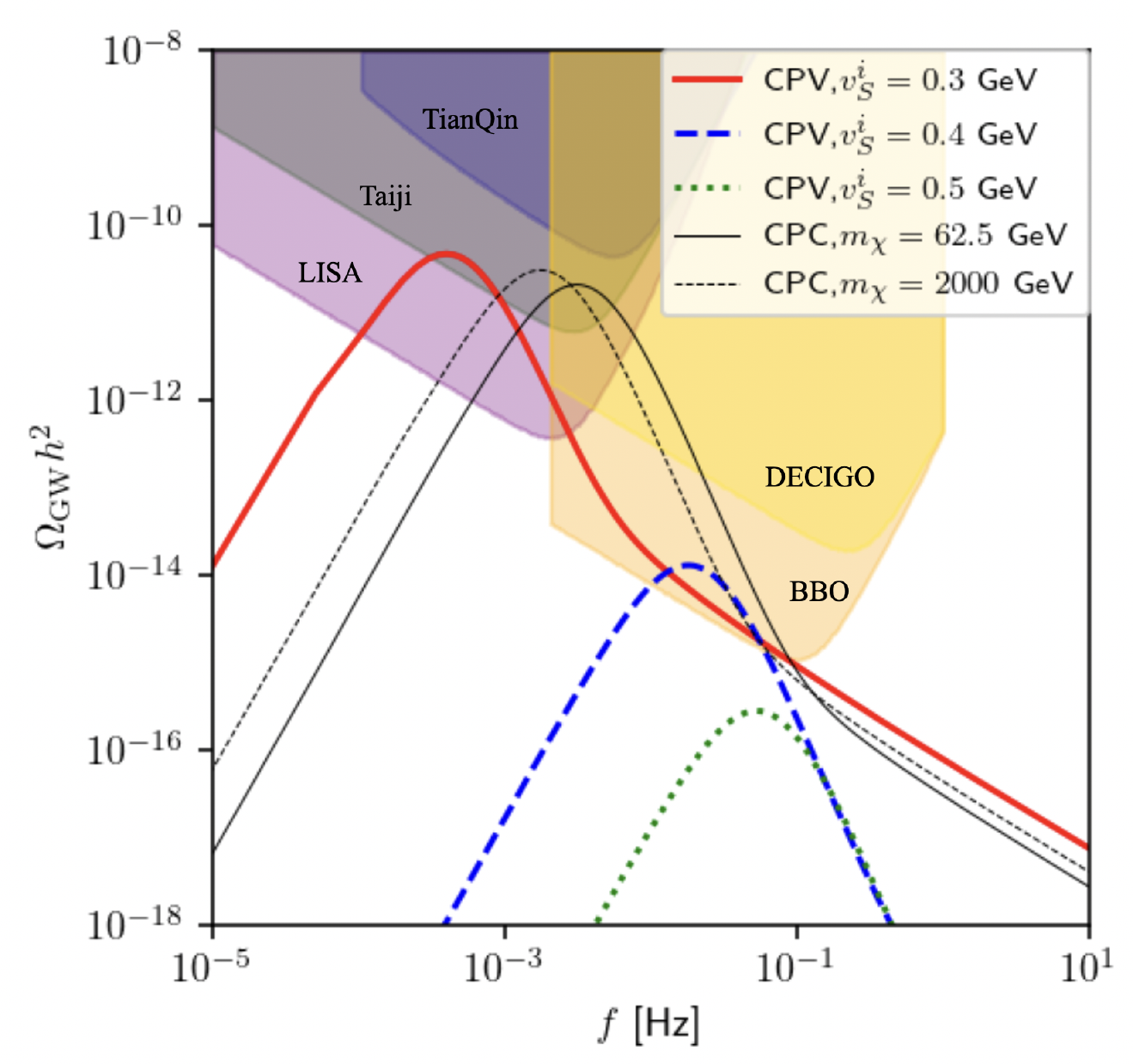}
\caption{$\Omega_{\text{GW}}h^2$ in the three BPs and two CPC cases are plotted as functions of $f$. 
The line and color schemes are coordinated as follows: BP1 (thick-solid, red), BP2 (thick-dashed, blue), BP3 (thick-dotted, green), CPC with $m_\chi=62.5$ GeV (thin-solid, black), and CPC with $m_\chi=2$ TeV (thin-dotted, black). The sensitivity curves of the future GW experiments TianQin, Taiji, LISA, DECIGO, and BBO are also shown. }
\label{fig:GW}
\end{figure}

In Fig.~\ref{fig:GW}, $\Omega_{\text{GW}}h^2$ in the three BPs and two CPC cases are shown. The line and color schemes are as follows:
BP1 (thick-solid, red), BP2 (thick-dashed, blue), BP3 (thick-dotted, green), CPC with $m_\chi=62.5$ GeV (thin-solid, black), and CPC with $m_\chi=2$ TeV (thin-dotted, black). We also overlay the sensitivity curves of the future GW experiments such as TianQin~\cite{TianQin:2015yph,Hu:2018yqb}, Taiji~\cite{Hu:2017mde,Ruan:2018tsw}, LISA~\cite{Caprini:2015zlo,amaro2017laser,Caprini:2019egz}, DECIGO~\cite{Seto:2001qf,Kawamura:2006up}, and BBO~\cite{Corbin:2005ny}. 
In BP1 and two CPC cases, $\Omega_{\text{GW}}h^2$ can be as large as $10^{-11}$ at the peak frequencies, which can be probed by Taiji, LISA, DECIGO, and BBO. On the other hand, it is found that $\Omega_{\text{GW}}h^2\simeq 10^{-14}$ in BP2 and  $\Omega_{\text{GW}}h^2\simeq 10^{-16}$ in BP3 at each peak frequency. 
Therefore, only BBO may be able to probe BP2, while BP3 is still beyond the reach of the proposed experiments.

Our numerical results show that although GW is not the CPV observable, it is highly sensitive to the size of CP violation in the singlet scalar sector since the strength of the first-order EWPT is primarily affected by it.
In the degenerate scalar scenario considered here, it is concluded that the GWs could be detectable for $\theta_S\lesssim 0.7$ radians.

Lastly, we address theoretical uncertainties in our calculation. It is known that the renormalization scale dependence on $\Omega_{\text{GW}}h^2$ could be significant if one uses the resummed one-loop effective potential (for a recent study, see, e.g., Ref.~\cite{Gould:2021oba}). Moreover, our calculation misses a finite lifetime factor of sound waves pointed out by Ref.~\cite{Guo:2020grp}. Due to those uncertainties, $\Omega_{\text{GW}}h^2$ could be modified by a factor to an order of magnitude. 
Nevertheless, our main conclusion would not be drastically changed. 

%------------------------------------------------------------------------------------------------------
%------------------------------------------------------------------------------------------------------

%%%%%%%%%%%%%%%%%%%%%%%%%%%%%%%%%%%%%%%%%%%%%
% Summary
%%%%%%%%%%%%%%%%%%%%%%%%%%%%%%%%%%%%%%%%%%%%%
\section{Summary and Discussions}\label{sec:sum}

We have investigated the possibility of a strong first-order EWPT taking CPV into account in the CxSM with degenerate scalars as a generalization of our previous study in Ref.~\cite{Cho:2021itv} where CP was assumed to be conserved in light of DM physics.
To understand the impacts of CP violation on the first-order EWPT qualitatively, we derived the analytical expressions of $v_C$ and $T_C$ [Eqs.~(\ref{vc}) and (\ref{Tc}) ] using the HT potential (tree-level potential together with the thermal masses). As is the CPC case, $v_C/T_C$ can be enhanced by the doublet-singlet Higgs mixing coupling $\delta_2$. However, we found that $\delta_2$ is related to the imaginary part of the singlet VEV $v_S^i$ [see Eq.~(\ref{del2})] in a way that $\delta_2$ decreases as $v_S^i$ increases. 
As a result, $v_C/T_C$ in the CPV case tends to be suppressed. This behavior was also numerically verified using the one-loop effective potential with the Parwani resummation scheme (Fig.~\ref{fig:EWPT_vsi}). 
Because of the simplicity of our scalar potential, the singlet phase $\theta_S$ is temperature independent, implying that the size of the CPV is fully determined by that at zero temperature, and there is no possibility of transitional CPV.
In the investigated parameter space, the least CPV case in BP1 gives more or less the similar $v_C/T_C$ as those in the CPC cases~\cite{Cho:2021itv} (Table~\ref{tab:atTc}).

We also evaluated the critical temperature and corresponding bubble wall profiles.
It was found that all of the bubble wall profiles have hyperbolic tangent forms and their widths become thinner as the first-order EWPT gets stronger (Fig.~\ref{fig:profile}), which is consistent with the known behavior. On the other hand, the phase of the singlet bubble wall is constant in space (Fig.~\ref{fig:ratio}), which is the reflection of the simple structure of the model setup. 

As an interesting consequence of the strong first-order EWPT, we also evaluated the GWs. 
Since GWs are modulated by the strength of strong first-order EWPT, the larger CPV in the singlet scalar can diminish the GW amplitudes significantly (Fig.~\ref{fig:GW}). 
In our degenerate scalar scenario, the signatures of the GWs may be seen in future experiments if $\theta_S \lesssim 0.7$ radians. Conversely, the larger CPV would not yield a detectable GW spectrum. 
In this way, even though the GW is not the CPV observable, we may get some useful information on CPV in the CxSM. We also emphasize that the GW signals would provide an exquisite probe of the degenerate scalars, which is currently the experimental blind spot. 

Besides the GW probes, detection of the interference effects among the degenerate scalars would be a smoking gun for our scenario.
For example, off-shell Higgs production cross sections, which can be used to extract the total Higgs decay width~\cite{Kauer:2012hd,Caola:2013yja}, might be sensitive to the interference effects. In addition, the degeneracy of the scalars could be disentangled by a recoil mass technique at lepton colliders~\cite{Abe:2021nih}. 
Therefore, future collider experiments such as High Luminosity-LHC and International Linear Collider would play a complementary role in probing the degenerate scalar scenario.

Finally, we make comments on theoretical issues in the current model.
As noticed, for successful EWBG, additional new particles are needed to induce CPV effects in the SM fermion sector. Furthermore, a candidate for DM should also be introduced (see, e.g., Ref.~\cite{Chen:2022vac}). Those issues will be addressed elsewhere.

\begin{acknowledgments}
We are grateful to Hiroto Shibuya for his valuable help with \texttt{CosmoTransitions} and helpful discussions. 
The work of G.C.C. is supported in part by JSPS KAKENHI Grant No. 22K03616. 

\end{acknowledgments}

%%%%%%%%%%%%%%%%%%%%%%%%%%%%%%%%%%%%%%%%%%%%%%%%
%							Appendix
%%%%%%%%%%%%%%%%%%%%%%%%%%%%%%%%%%%%%%%%%%%%%%%%
\appendix

\section{Field-dependent masses and thermal masses}\label{app:fm}
Here we list the field-dependent masses and derive the thermal masses of the scalars. 
In the presence of CP violation, the mass matrix of the scalars takes the 3-by-3 form
\begin{align}
\frac{1}{2}
\begin{pmatrix}
\varphi & \varphi_S^r & \varphi_S^i 
\end{pmatrix}
\bar{\mathcal{M}}_S^2
\begin{pmatrix}
\varphi \\ 
\varphi_S^r \\
\varphi_S^i 
\end{pmatrix},
\end{align}
where
\begin{align}
(\bar{\mathcal{M}}_S^2)_{11}&=\frac{m^2}{2}+\frac{3\lambda}{4}\varphi^2+\frac{\delta_2}{4}|\varphi_S|^2,\\
(\bar{\mathcal{M}}_S^2)_{22}&=\frac{b_2}{2}+\frac{b_1^r}{2}+\frac{d_2}{4}(3\varphi_S^{r2}+\varphi_S^{i2})+\frac{\delta_2}{4}\varphi^2,\\
(\bar{\mathcal{M}}_S^2)_{33}&= \frac{b_2}{2}-\frac{b_1^r}{2}+\frac{d_2}{4}(\varphi_S^{r2}+3\varphi_S^{i2})+\frac{\delta_2}{4}\varphi^2,\\
(\bar{\mathcal{M}}_S^2)_{12}&= \frac{\delta_2}{2}\varphi\varphi_S^r,\\
(\bar{\mathcal{M}}_S^2)_{13}&= \frac{\delta_2}{2}\varphi\varphi_S^i,\\
(\bar{\mathcal{M}}_S^2)_{23}&= -\frac{b_1^i}{2}+\frac{d_2}{2}\varphi_S^r\varphi_S^i,
\end{align}
with $|\varphi_S|^2=\varphi_S^{r2}+\varphi_S^{i2}$.
The field-dependent masses of the NG bosons, gauge bosons and top/bottom are, respectively, given by
\begin{align}
\bar{m}_{G^0}^2&= \bar{m}_{G^\pm}^2
= \frac{m^2}{2}+\frac{\lambda}{4}\varphi^2
	+\frac{\delta_2}{4}\varphi_S^2, \\
\bar{m}_{W}^2&=\frac{g_2^2}{4}\varphi^2, \quad
\bar{m}_{Z}^2 =\frac{g_2^2+g_1^2}{4}\varphi^2, \\
\bar{m}_t^2 &= \frac{y_t^2}{2}\varphi^2,\quad \bar{m}_b^2 = \frac{y_b^2}{2}\varphi^2.
\end{align}
The thermal masses of $\varphi$ to $\mathcal{O}(T^2)$ appearing in the potential (\ref{VHT}) is calculated as follows. 
\begin{align}
\Sigma_HT^2 &=  \frac{\partial^2 V_1|_{T>0}}{\partial \varphi^2} = \sum_in_i
\left[
\frac{T^2}{2\pi^2}\frac{\partial I_{B,F}(a_i^2)}{\partial a_i^2}\frac{\partial^2 \bar{m}_i^2}{\partial\varphi^2}
+\frac{1}{2\pi^2}\frac{\partial^2 I_{B,F}(a_i^2)}{\partial (a_i^2)^2}\left(\frac{\partial \bar{m}_i^2}{\partial\varphi}\right)^2
\right] \nonumber\\
&\simeq \frac{T^2}{24}\sum_{i=\text{bosons}}n_i\frac{\partial^2 \bar{m}_i^2}{\partial\varphi^2}-
\frac{T^2}{48}\sum_{i=\text{fermions}}n_i\frac{\partial^2 \bar{m}_i^2}{\partial\varphi^2} \nonumber\\
%& = \frac{T^2}{24}\frac{\partial^2}{\partial\varphi^2}
%(\bar{m}_{h_1}^2+\bar{m}_{h_2}^2+\bar{m}_{h_3}^2+\bar{m}_{G^0}^2+2\bar{m}_{G^\pm}^2
%+6\bar{m}_{W}^2+3\bar{m}_{Z}^2)-\frac{T^2}{48}(-12)\frac{\partial^2}{\partial\varphi^2}(\bar{m}_t^2+\bar{m}_b^2) \nonumber\\
& = \left[\frac{\lambda}{8}+\frac{\delta_2}{24}+\frac{1}{16}(3g_2^2+g_1^2)+\frac{1}{4}(y_t^2+y_b^2)\right]T^2,
\end{align}
where $\sum_{i=1,2,3}\bar{m}_i^2 = \sum_{i=1,2,3}(\bar{\mathcal{M}}_S^2)_{ii}$ is used for the scalars. 
Similarly, the thermal masses of the singlet fields $\varphi_S^r$ and $\varphi_S^i$ are found to be
\begin{align}
\Sigma_ST^2 &= \frac{\partial^2 V_1|_{T>0}}{\partial \varphi_S^{r2}} = \frac{\partial^2 V_1|_{T>0}}{\partial \varphi_S^{i2}} = \frac{\delta_2+d_2}{12}T^2. 
\end{align}

%%%%%%%%%%%%%%%%%%%%%%%%%%%%%%%%%%%%%%%%%%%%%%%%
%							References
%%%%%%%%%%%%%%%%%%%%%%%%%%%%%%%%%%%%%%%%%%%%%%%%

\bibliography{refs}
\nocite{Schicho:2022wty}

\end{document}